\documentclass[sigconf]{acmart}
\usepackage{multirow}
\usepackage{balance}
\usepackage[ruled,vlined,linesnumbered]{algorithm2e}
\AtBeginDocument{%
  }

\setcopyright{acmlicensed}
\copyrightyear{2018}
\acmYear{2018}
\acmDOI{XXXXXXX.XXXXXXX}
\acmConference[Conference acronym 'XX]{Make sure to enter the correct
  conference title from your rights confirmation email}{June 03--05,
  2018}{Woodstock, NY}

\acmISBN{978-1-4503-XXXX-X/2018/06}

\begin{document}

\title{MIMA: Multi-Interest Recommendation via Multi-Positive Exclusive Assignment}

\author{Xingyuan Mao}
\affiliation{%
  \institution{Alibaba International Digital Commerce Group}
  \city{Beijing}
  \country{China}}
\email{maoxingyuan.mxy@alibaba-inc.com}

\author{Alin Fan}
\affiliation{%
  \institution{Alibaba International Digital Commerce Group}
  \city{Beijing}
  \country{China}}
\email{alin.fal@alibaba-inc.com}

\author{Shichao Nie}
\affiliation{%
  \institution{Alibaba International Digital Commerce Group}
  \city{Guangzhou}
  \country{China}}
\email{nsc383441@alibaba-inc.com}

\author{Junfeng Zhang}
\affiliation{%
  \institution{Alibaba International Digital Commerce Group}
  \city{Guangzhou}
  \country{China}}
\email{sichu.zjf@alibaba-inc.com}

\author{Yan Xiao}
\affiliation{%
  \institution{Alibaba International Digital Commerce Group}
  \city{Guangzhou}
  \country{China}}
\email{yanwei.xy@alibaba-inc.com}

\author{Tao Luo}
\affiliation{%
  \institution{Alibaba International Digital Commerce Group}
  \city{Beijing}
  \country{China}}
\email{luotao.lt@alibaba-inc.com}

\author{Xiaoyi Zeng}
\affiliation{%
  \institution{Alibaba International Digital Commerce Group}
  \city{Hangzhou}
  \country{China}}
\email{yuanhan@alibaba-inc.com}

\renewcommand{\shortauthors}{Mao et al.}


\begin{abstract}
Multi-interest recommendation represents each user with multiple interest vectors for fine-grained candidate matching, yet it often suffers from interest collapse, where the learned interests converge to similar representations. We highlight the prevailing single-positive paradigm as one important factor behind this issue. Since each instance provides only one positive item, intents are optimized independently, potentially causing the same best-matching interest to be repeatedly updated toward different positives while leaving the others under-supervised. Moreover, existing methods rarely model how strongly a user activates each interest, leaving scores from different interest channels incomparable at inference. To address these problems, we propose MIMA, a \underline{M}ulti-\underline{I}nterest recommendation framework built on \underline{M}ulti-positive exclusive \underline{A}ssignment. MIMA groups items co-occurring within the same request into a positive set, generates complementary interests with a causal Transformer decoder, and exclusively assigns each positive to supervise a distinct interest via Hungarian matching, so that interest differentiation emerges from the training objective itself rather than auxiliary regularization. A lightweight routing module further estimates user-interest activation probabilities to calibrate scores across interest channels. Experiments on three public datasets and an industrial dataset show that MIMA consistently outperforms state-of-the-art baselines, and an online A/B test yields significant business gains.
\end{abstract}


\maketitle
\section{Introduction}

Modern recommender systems rely on effective user representations to capture preferences and retrieve relevant items~\cite{covington2016deep, lv2019sdm, liu2025facet}. Traditional methods usually represent each user with a single embedding vector, limiting their ability to capture diverse and dynamic user interests~\cite{kang2018self, sun2019bert4rec}. In recent years, multi-interest recommendation has become a key technique in industrial-scale recommender systems~\cite{li2019multi,cen2020controllable, tian2022multi}. It represents each user with multiple interest vectors, each capturing a distinct preference facet~\cite{li2026multi}, and supports fine-grained candidate matching through approximate nearest neighbor search over multiple interest channels~\cite{johnson2019billion,chai2022user}.

Despite this promise, multi-interest recommendation still suffers from the \textbf{interest collapse} problem, where the learned interest vectors converge to similar representations rather than specialize in distinct user intents~\cite{zhang2022re4,xie2023rethinking,du2024disentangled}. This collapse weakens the expressive ability of multi-interest models, as redundant interest vectors retrieve highly overlapping candidates and fail to cover users' diverse preferences. Existing studies alleviate this problem mainly through auxiliary constraints that encourage diversity~\cite{zhang2022re4}, balance routing~\cite{xie2023rethinking, lee2024towards}, or guide disentangled item-interest assignment~\cite{du2024disentangled}. Although effective to varying degrees, these constraints lie outside the primary recommendation objective and thus exert only indirect pressure for interest specialization. This motivates us to look beyond auxiliary constraints and revisit the supervision paradigm shared by existing methods: \textit{single-positive} training.

\begin{figure}[t]
    \centering
    \includegraphics[width=\linewidth]{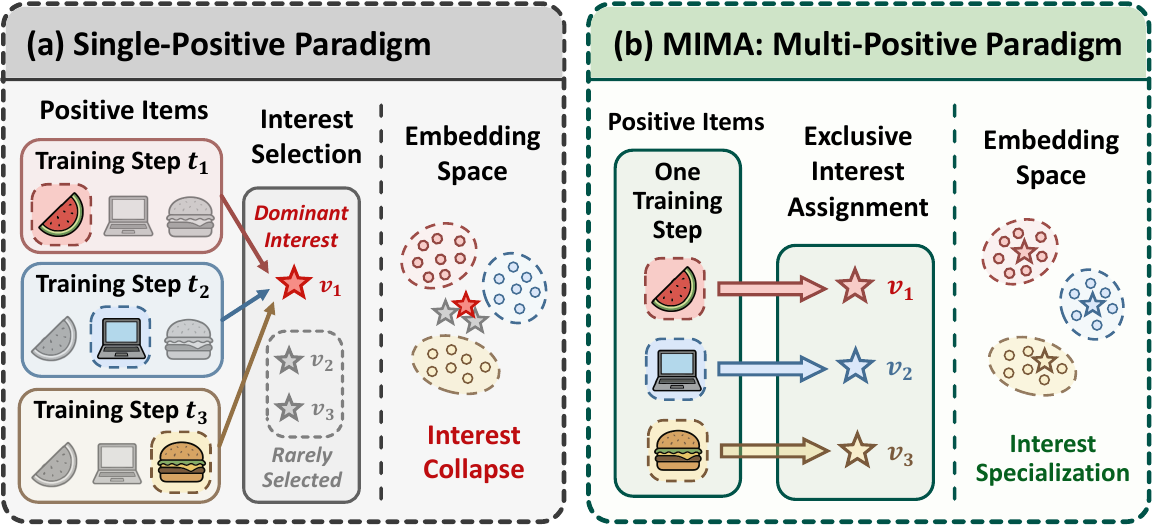}

    \caption{Comparison between (a) the single-positive paradigm and (b) our multi-positive paradigm. Positive items are drawn from the interaction sequence of the same user. While single-positive training routes isolated positives to a dominant interest and causes interest collapse, MIMA makes concurrent positives jointly visible and exclusively assigns each to a distinct interest, enabling interest specialization.}
    \label{fig:intro}
\end{figure}

Under the single-positive paradigm, each training instance contains only one positive item, and the model matches it with one of the learned interest vectors. However, this isolated formulation provides limited supervision for how multiple interests should specialize and separate. In real user behavior, multiple items clicked within the same request often reflect concurrent but distinct intents~\cite{hu2019sets2sets,li2023next,guan2026make}. Ideally, a multi-interest model should encourage different interest vectors to take primary responsibility for different concurrent positives, so that each vector can specialize in a distinct preference facet~\cite{li2023multiintention,liu2025facet}. However, single-positive training separates these positives into different training instances, making them mutually invisible during each optimization step. Without observing these positives jointly, the model cannot learn explicit competition among interest vectors. Instead, as illustrated in Figure~\ref{fig:intro}(a), each isolated positive is routed to the currently best-matching interest, so a dominant interest repeatedly wins across steps and absorbs conflicting gradients from distinct intents, gradually drifting toward an averaged direction, while the remaining interests are rarely selected, receive no distinct supervision and merely drift with the dominant interest through shared parameters.

While interest collapse arises during training, another problem emerges at inference time when candidates from different interest channels are merged~\cite{li2019multi,cen2020controllable,meng2025user}. We refer to this issue as \textbf{cross-interest score incomparability}.
Existing methods mainly model the matching between an interest vector and an item, but usually do not explicitly model how likely each interest is to be activated for the current user~\cite{chai2022user,wang2022target}. During inference, candidate items retrieved from different interest channels are merged and ranked together. However, scores from different interest channels lack a unified normalization basis, since a high matching score from a weakly activated interest may be treated as comparable to a score from a dominant interest. This mismatch makes cross-interest ranking unreliable and may lead to suboptimal retrieval results~\cite{steck2018calibrated}.

To address these limitations, we propose \textbf{MIMA}, a \textbf{M}ulti-\textbf{I}nterest recommendation framework based on \textbf{M}ulti-positive exclusive \textbf{A}ssignment. Instead of constructing each training instance with a single positive item, MIMA groups co-occurring items within the same request into a positive set, making concurrent user intents jointly visible in the same training step. Given this multi-positive supervision, a causal Transformer decoder~\cite{vaswani2017attention} first generates complementary interests, where each interest query attends only to earlier queries, encouraging later interests to complement earlier ones and yielding differentiated representations. Based on these complementary interests, we further introduce an interest exclusivity mechanism during training, which applies Hungarian optimal assignment~\cite{kuhn1955hungarian,carion2020end} between the interest vectors and multiple positive items. As illustrated in Figure~\ref{fig:intro}(b), this exclusive one-to-one matching enables interests to naturally compete, so that different positives route their gradients to different interest vectors, promoting specialization and mitigating interest collapse. To further address cross-interest score incomparability at inference, a lightweight routing network estimates user-interest activation probabilities, which are then combined with interest-item matching scores to obtain calibrated retrieval scores across interest channels. In this way, MIMA models each user-item relationship by jointly considering how well an item matches a specific interest and how strongly that interest is activated for the user. 

Our contributions are summarized as follows:

\begin{itemize}
    \item We propose MIMA, to the best of our knowledge, the first work to introduce the multi-positive paradigm into multi-interest recommendation. By generating complementary interests and assigning them exclusively to multiple positive items, MIMA drives different interests toward distinct positive signals, thereby mitigating interest collapse.

    \item We introduce user-interest routing to model how strongly each interest is activated for a user, mitigating score incomparability across multiple interest channels.

    \item We conduct extensive experiments on three public datasets and a large-scale industrial dataset, followed by an online A/B test in production. The results show that MIMA consistently outperforms existing methods.
\end{itemize}

\section{Related Work}

\subsection{Candidate Matching}
Candidate matching retrieves relevant items from a large corpus and serves as the foundation for subsequent ranking in industrial recommender systems. Early studies mainly rely on collaborative filtering or matrix factorization to learn user and item representations from implicit feedback~\cite{sarwar2001item,koren2009matrix,hu2008collaborative}. With the development of deep learning, two-tower models such as DSSM~\cite{huang2013learning} and YouTube-DNN~\cite{covington2016deep} become a common retrieval paradigm, where one tower encodes the user and the other encodes candidate items into a shared embedding space for efficient nearest-neighbor search. Subsequent works extend this paradigm by incorporating sequential user behaviors~\cite{lv2019sdm} or replacing the fixed inner product with neural functions~\cite{he2017neural}. Recent studies further apply graph neural networks to capture high-order collaborative signals~\cite{wang2019neural,he2020lightgcn,zhang2024linear}. Despite their effectiveness, these methods typically represent each user with a single vector, making it difficult to capture diverse user interests.

\subsection{Multi-Interest Recommendation}
Multi-interest recommendation represents a user with multiple vectors to capture diverse preference facets~\cite{li2026multi, lee2024towards}. MIND~\cite{li2019multi} pioneers this paradigm by clustering historical items into interest capsules via dynamic routing~\cite{sabour2017dynamic}, while ComiRec~\cite{cen2020controllable} further introduces a self-attention variant. Later studies enrich interest extraction with temporal periodicity~\cite{chen2021exploring}, user profiles~\cite{chai2022user}, and target-interest distillation~\cite{wang2022target}. Recent methods further explore uncertainty-aware neural processes~\cite{jiang2025auto} and generative interest quantization~\cite{wu2025gemirec}. Another line of work focuses on mitigating interest collapse~\cite{zhang2022re4,xie2023rethinking,du2024disentangled}. REMI~\cite{xie2023rethinking} adopts interest-aware hard negative mining and routing regularization. DisMIR~\cite{du2024disentangled} leverages item co-occurrence graphs for disentangled interest learning. However, existing methods still face two limitations. First, they mitigate interest collapse only through auxiliary regularization or constraints outside the primary training objective. Second, they model only interest--item matching scores without explicit user--interest activation, leaving scores across interest channels incomparable.

\subsection{Multi-Positive Learning}
Multi-positive learning exploits multiple positive items or labels observed under the same user context as training signals. This idea relates to multi-label learning~\cite{zhang2014review}, where one input is associated with multiple relevant labels, and connects to recommendation, tagging, and ranking tasks~\cite{jain2016extreme,liu2017deep}. In recommendation, next-basket recommendation~\cite{yu2016dynamic,li2023next} and sequential set prediction~\cite{hu2019sets2sets} naturally treat multiple future items as positives. Multi-positive contrastive learning further shows that multiple positives provide richer supervision than a single positive pair~\cite{khosla2020supervised,li2023multiintention}. However, existing studies use multiple positives mainly for label prediction, basket prediction, or contrastive learning, leaving their role in driving differentiated multi-interest representations underexplored.

\section{Preliminaries}
\subsection{Problem Formulation}
Given a user set $\mathcal{U}$ and a large item corpus $\mathcal{I}$, each user $u\in\mathcal{U}$ has a chronological behavior sequence $s_u=(i^u_1,i^u_2,\ldots,i^u_n)$, where $i^u_m\in\mathcal{I}$ denotes the $m$-th interacted item and $n$ is the maximum sequence length. The candidate matching stage aims to efficiently retrieve a subset of items that the user is likely to interact with from $\mathcal{I}$. In multi-interest recommendation, the model represents user $u$ with an interest matrix $\mathbf{V}_u=[\mathbf{v}^u_1,\mathbf{v}^u_2,\ldots,\mathbf{v}^u_K]\in\mathbb{R}^{K\times d}$, where $\mathbf{v}^u_k\in\mathbb{R}^d$ denotes the $k$-th interest vector of user $u$ and $d$ is the embedding dimension. These interest vectors are then used to perform multiple retrievals over the item corpus, and the retrieved candidates are merged as the output of the matching stage.

\subsection{Single-Positive Paradigm}
Most existing multi-interest recommendation methods follow a single-positive training paradigm~\cite{li2019multi,cen2020controllable,du2024disentangled}. Given a user's behavior sequence $s_u$, the model encodes the sequence into $K$ interest vectors:
\begin{equation}
    \mathbf{V}_u=\mathcal{F}(s_u)=[\mathbf{v}^u_1,\mathbf{v}^u_2,\ldots,\mathbf{v}^u_K],
\end{equation}
where $\mathcal{F}(\cdot)$ is a multi-interest extractor, commonly implemented with attention mechanisms~\cite{cen2020controllable} or capsule networks~\cite{li2019multi,du2024disentangled}.

During training, each instance is constructed as a pair $(s_u,y)$, where $y$ is the only positive item. The model determines which specific interest vector should be responsible for predicting $y$. This is typically achieved through a hard-routing mechanism that selects the most aligned interest based on the inner product between the generated interest vectors and the positive item's embedding $\mathbf{e}_y$. The index of the activated interest vector is determined by:
\begin{equation}
    \label{eq:hard-routing}
    k^*=\arg\max_{1\le k\le K}(\mathbf{v}^u_k)^\top\mathbf{e}_y.
\end{equation}

Subsequently, the model uses this maximally activated interest vector to compute the final matching score for the positive item. The predicted preference score is the inner product of the selected interest and the positive item's embedding:
\begin{equation}
    \hat{r}_{u,y}=(\mathbf{v}^u_{k^*})^\top\mathbf{e}_y.
\end{equation}

Although this paradigm is widely adopted, it provides only one positive item in each optimization step. Multiple positives that occur under the same user context are split into separate training instances and become mutually invisible during a single forward pass. Without joint visibility, these positives cannot compete for distinct interest vectors. Each update only pulls the single selected interest, typically the dominant one, toward an isolated positive, leaving other interests scarcely supervised and making redundant interest representations a stable solution of training.

\section{Methodology}

\begin{figure*}[t]
    \centering
    \includegraphics[width=0.97\textwidth]{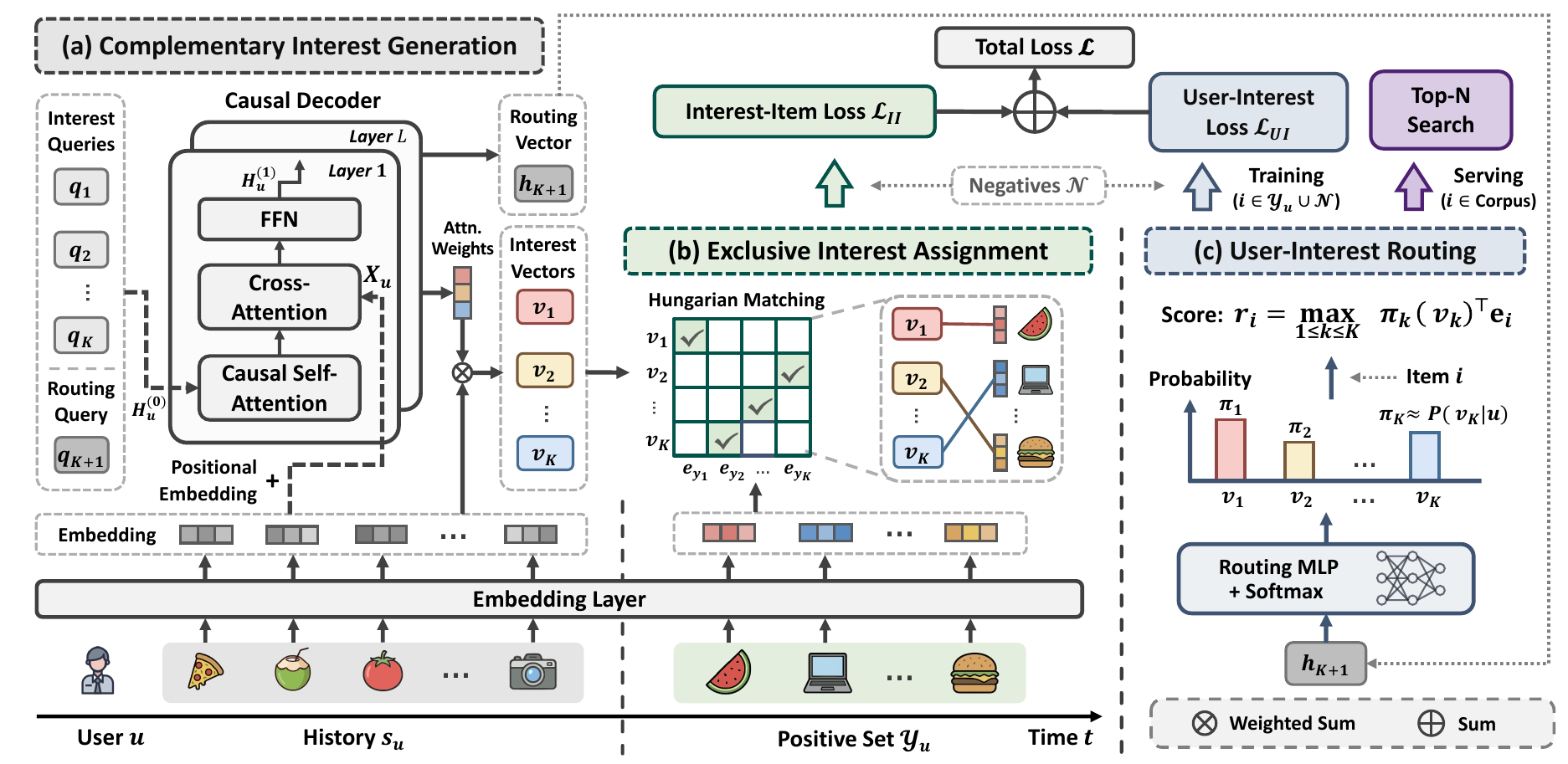}

    \caption{The overall framework of MIMA. Items co-occurring within the same request are grouped into a positive set. Then (a) a causal decoder generates complementary interests, (b) Hungarian matching exclusively assigns each positive to a distinct interest, and (c) a routing module estimates user-interest activation probabilities to calibrate per-interest scores.}
    \label{fig:method}
\end{figure*}

This section details the MIMA framework, which operates under a multi-positive paradigm. As illustrated in Figure~\ref{fig:method}, MIMA first constructs positive sets by grouping concurrent behaviors. Next, a causal decoder generates complementary interest vectors, which are then forced to compete for distinct positive items via exclusive assignment. Finally, a user-interest routing module calibrates retrieval scores across different interest channels.

\subsection{Multi-Positive Target Construction}
\label{sec:multi-positive}
The first step toward multi-positive supervision is to construct set-level positive targets. In industrial recommender systems, each request exposes a page of candidate items, of which the user may click several, making the request a natural grouping unit~\cite{guan2026make,song2026tmallgs}. Since items clicked within the same request share exactly the same user-side features and thus form strictly concurrent positives, we group them into one positive set per request.

Formally, given user $u$'s chronologically sorted interactions, let $i_m^u$ be the $m$-th interacted item and $\rho_m^u$ the index of the request under which it occurs. We construct a positive set for the $b$-th request as:
\begin{equation}
\label{eq:pos-set}
\mathcal{Y}_u^{(b)}=\operatorname{Unique}_K\!\left([\,i_m^u\mid \rho_m^u=b\,]\right),
\end{equation}
where $\operatorname{Unique}_K(\cdot)$ removes duplicated items and keeps the first $K$ unique positives, matching the set capacity to the interest number for the later exclusive assignment. When fewer than $K$ valid positives are available, we pad $\mathcal{Y}_u^{(b)}$ to a fixed length $K$ with a dummy item whose embedding is fixed to zero and excluded from gradient updates, and distinguish real positives from padded positions with a binary mask vector $\mathbf{m}_u^{(b)} \in \{0, 1\}^K$, whose $j$-th element is:
\begin{equation}
    \label{eq:pos-mask}
    m_{u,j}^{(b)} = \begin{cases} 
      1, & \text{if } j \le |\mathcal{Y}_u^{(b)}|, \\
      0, & \text{otherwise},
   \end{cases}
\end{equation}
where $|\mathcal{Y}_u^{(b)}|$ denotes the number of valid positive items before padding.
To avoid data leakage, each positive set $\mathcal{Y}_u^{(b)}$ is paired with the user's historical sequence available before the $b$-th request:
\begin{equation}
    \label{eq:hist-seq}
    s_u^{(b)} = \left( i_m^u \mid \rho_m^u < b \right).
\end{equation}

In this way, the supervision unit is upgraded from a single target $(s_u,y)$ to a set-level target $(s_u^{(b)},\mathcal{Y}_u^{(b)})$, where jointly visible positives can compete for distinct interest vectors, laying the foundation for the exclusive assignment below.

\subsection{Complementary Interest Generation}
\label{sec:interest-generation}
Multi-positive targets determine what the model should predict, but not how the interests themselves should be formed. Standard multi-interest extractors, whether built on attention mechanisms~\cite{cen2020controllable} or capsule networks~\cite{li2019multi}, derive all interest vectors independently from the same history, so nothing prevents different interests from locking onto similar behavior patterns.
To make each interest capture a distinct pattern, we propose Complementary Interest Generation, a causal Transformer decoder where each later query conditions on previously formed interest states and focuses on patterns not yet captured.
For notation simplicity, we omit the request index $(b)$ below. We map each historical item $i_m^u$ to an embedding $\mathbf{e}_{i_m^u}\in\mathbb{R}^d$ through the embedding table. A positional embedding $\mathbf{p}_m$ is then added to construct the historical representation matrix:
\begin{equation}
\label{eq:hist-repr}
\mathbf{X}_u=[\mathbf{e}_{i_1^u}+\mathbf{p}_1,\ldots,\mathbf{e}_{i_n^u}+\mathbf{p}_n]^\top\in\mathbb{R}^{n\times d}.
\end{equation}

MIMA maintains $K+1$ learnable queries:
\begin{equation}
\mathbf{Q}=[\mathbf{q}_1,\ldots,\mathbf{q}_K,\mathbf{q}_{K+1}]\in\mathbb{R}^{(K+1)\times d}.
\end{equation}

The first $K$ queries are interest queries, whereas $\mathbf{q}_{K+1}$ is a routing query excluded from interest generation. Its output is reserved for the user-interest routing module discussed in Section~\ref{sec:user-interest-routing}.

At each decoder layer, the queries first interact through causal self-attention and then attend to the historical representations through cross-attention before being processed by a feed-forward network. Starting from $\mathbf{H}_u^{(0)}=\mathbf{Q}$, the decoder performs the following updates at layer $\ell$:
\begin{equation}
\label{eq:decoder-layer}
\begin{gathered}
\widetilde{\mathbf{H}}_u^{(\ell)}=\operatorname{CausalSelfAttn}_{\ell}\left(\mathbf{H}_u^{(\ell-1)}\right),\\
\left(\overline{\mathbf{H}}_u^{(\ell)},\mathbf{A}_u^{(\ell)}\right)=\operatorname{CrossAttn}_{\ell}\left(\widetilde{\mathbf{H}}_u^{(\ell)},\mathbf{X}_u\right),\\
\mathbf{H}_u^{(\ell)}=\operatorname{FFN}_{\ell}\left(\overline{\mathbf{H}}_u^{(\ell)}\right).
\end{gathered}
\end{equation}

Here, $\widetilde{\mathbf{H}}_u^{(\ell)}$ and $\overline{\mathbf{H}}_u^{(\ell)}$ denote the intermediate query states. After $L$ layers, the decoder produces the final query states $\mathbf{H}_u=\mathbf{H}_u^{(L)}=[\mathbf{h}_1^u,\ldots,\mathbf{h}_{K+1}^u]\in\mathbb{R}^{(K+1)\times d}$ and the final-layer cross-attention matrix $\mathbf{A}_u=\mathbf{A}_u^{(L)}\in\mathbb{R}^{(K+1)\times n}$, where $A_{k,m}^u$ denotes the attention weight assigned by the $k$-th query to the $m$-th historical item.

Owing to the causal self-attention, the query state fed into cross-attention by the $k$-th query depends only on $\mathbf{q}_{1:k}$ and $\mathbf{X}_u$, so the attention distribution $\mathbf{A}_u[k,:]$ accounts for the interest states formed by preceding queries. This encourages later queries to attend to behavior patterns not yet captured, driving the interests toward complementary rather than redundant representations. Rather than directly using $\mathbf{h}_k^u$ as an interest vector, we aggregate historical item embeddings with the first $K$ rows of $\mathbf{A}_u$, keeping the resulting interests in the same representation space as candidate items:
\begin{equation}
\label{eq:interest-agg}
\mathbf{v}_k^u=\sum_{m=1}^{n}A_{k,m}^u\mathbf{e}_{i_m^u}.
\end{equation}

This yields the multi-interest representation $\mathbf{V}_u=[\mathbf{v}_1^u,\ldots,\mathbf{v}_K^u]$, and these interests are encouraged to capture distinct behavior patterns. However, the causal decoder only promotes complementarity among the interests and does not determine which interest should be responsible for each positive item. Therefore, to convert the previously constructed positive sets into differentiated supervision, we next introduce an explicit assignment mechanism that maps different positives to distinct interests.

\subsection{Exclusive Interest Assignment}
Given the interests $\mathbf{V}_u$ and the positive set $\mathcal{Y}_u$, we need to assign positive items to interests during training. Standard hard routing in Eq.~\eqref{eq:hard-routing} is designed for the single-positive paradigm, routing the sole positive to the interest with the largest inner product. However, when applied to multiple positives, it may assign them to the same dominant interest and leave others unsupervised, making it inadequate for our multi-positive paradigm. We therefore propose an exclusive assignment mapping each positive to a distinct interest, formalized as a linear assignment problem and solved exactly.

We first compute the similarity matrix $\mathbf{C}_u\in\mathbb{R}^{K\times K}$ between interests and positives, whose entries are inner products:
\begin{equation}
\label{eq:sim-matrix}
C_{k,j}^u=(\mathbf{v}_k^u)^\top\mathbf{e}_{y_j},
\end{equation}
where $y_j$ denotes the $j$-th positive item in $\mathcal{Y}_u$. We then introduce a binary assignment matrix $\mathbf{Z}\in\{0,1\}^{K\times K}$, where $Z_{k,j}=1$ indicates that positive $y_j$ is assigned to interest $\mathbf{v}_k^u$. The exclusive assignment is defined as the solution of the following integer program:
\begin{equation}
\label{eq:exclusive-assignment}
\begin{gathered}
\mathbf{Z}^{*}=\arg\max_{\mathbf{Z}\in\{0,1\}^{K\times K}}\ \textstyle\sum_{k=1}^{K}\sum_{j=1}^{K}Z_{k,j}C_{k,j}^u\\
\mathrm{s.t.}\ \textstyle\sum_{k}Z_{k,j}=m_{u,j},\ \forall\,j;\quad \sum_{j}Z_{k,j}\le 1,\ \forall\,k.
\end{gathered}
\end{equation}

The two constraints enforce bidirectional exclusivity: the equality constraint guarantees that every valid positive is assigned to exactly one interest, where the binary mask $m_{u,j}$ in Eq.~\eqref{eq:pos-mask} excludes padded positions from the matching, while the inequality constraint ensures that each interest supervises at most one positive.

Eq.~\eqref{eq:exclusive-assignment} is a standard linear assignment problem. By negating the similarity matrix as the cost matrix, it can be solved exactly by the Hungarian algorithm~\cite{kuhn1955hungarian}. The algorithm iteratively performs row and column reductions, seeks a maximum set of independent zero entries, and adjusts the matrix with the smallest uncovered element until these zeros define the optimal one-to-one matching. Given the optimal assignment $\mathbf{Z}^{*}$, the interest vector responsible for the $j$-th positive $y_j$ is read out as:
\begin{equation}
\label{eq:assigned-interest}
\bar{\mathbf{v}}_j^u=\sum_{k=1}^{K}Z^{*}_{k,j}\mathbf{v}_k^u,
\end{equation}
which is supervised by positive $y_j$ in the training objective, while unmatched interests receive no gradient. The assignment is performed under stop-gradient. $\mathbf{Z}^{*}$ only selects the interest-positive pairs participating in the loss, and gradients flow through the loss back to the selected interests and the decoder.

It is worth emphasizing that the exclusive assignment is an integral part of the multi-positive paradigm rather than an add-on. By routing different positives to distinct interests, it prevents a dominant interest from absorbing multiple positive signals within the same instance, so interest differentiation emerges from the training objective itself, intrinsically alleviating interest collapse.

Besides the Hungarian algorithm, we also explore two alternatives: a basic greedy algorithm that sequentially assigns each positive to its highest-scoring interest among the remaining ones, and a Sinkhorn-based soft assignment~\cite{sinkhorn1964relationship,cuturi2013sinkhorn}. The greedy solution is order-dependent and only locally optimal, while Sinkhorn produces only a soft assignment that is approximate and non-exclusive. In contrast, the Hungarian assignment achieves exact global optimality and order independence. Experiments in Section~\ref{sec:assignment-analysis} confirm that the Hungarian assignment performs best overall.

\subsection{User-Interest Routing}
\label{sec:user-interest-routing}
Having obtained differentiated interests, the remaining task is to merge the $K$ per-channel scores $(\mathbf{v}_k^u)^\top\mathbf{e}_i$ into a single ranking for retrieval. To see what a principled merge requires, we take a probabilistic view and factorize the preference of user $u$ for item $i$ as:
\begin{equation}
\label{eq:pref-factorization}
P(i\mid u)=\sum_{k=1}^{K}P(i\mid\mathbf{v}_k^u)\,P(\mathbf{v}_k^u\mid u),
\end{equation}
where $P(i\mid\mathbf{v}_k^u)$ measures the interest-item relevance and $P(\mathbf{v}_k^u\mid u)$ reflects the strength with which user $u$ activates the $k$-th interest. Existing methods model only the former via inner products while implicitly treating the latter as uniform, so scores from different channels lack a shared calibration basis and are not directly comparable. For example, an item scoring 0.8 under a marginal interest that the user rarely activates would outrank an item scoring 0.7 under the user's dominant interest, even though the latter is more likely to be interacted with. Such incomparable scores distort the merged ranking in the matching stage. We therefore introduce a lightweight routing module that explicitly estimates the activation distribution $P(\mathbf{v}_k^u\mid u)$ over the $K$ interests.

We use the routing query $\mathbf{q}_{K+1}$ reserved in Section~\ref{sec:interest-generation}. Owing to causal self-attention, its final state $\mathbf{h}_{K+1}^u$ has attended to the states of all $K$ interest queries and thus serves as a global summary of the user's interests at no extra encoding cost. The routing distribution is obtained by a two-layer MLP:
\begin{equation}
\label{eq:routing}
\boldsymbol{\pi}_u=\operatorname{softmax}\left(\mathbf{W}_2\,\sigma\left(\mathbf{W}_1\operatorname{sg}\left[\mathbf{h}_{K+1}^u\right]\right)\right),
\end{equation}
where $\mathbf{W}_1\in\mathbb{R}^{d\times d}$ and $\mathbf{W}_2\in\mathbb{R}^{K\times d}$ are weight matrices, $\sigma(\cdot)$ is the LeakyReLU activation, and $\operatorname{sg}[\cdot]$ denotes the stop-gradient operator. The resulting $\boldsymbol{\pi}_u=[\pi_1^u,\ldots,\pi_K^u]$ estimates the activation distribution, i.e., $\pi_k^u\approx P(\mathbf{v}_k^u\mid u)$. The stop-gradient isolates routing from representation learning, so the routing objective only updates the MLP, and the interest and item representations remain exclusively shaped by the interest-item loss in Eq.~\eqref{eq:interest-item-loss}.

With $\boldsymbol{\pi}_u$, we define the calibrated joint score between user $u$ and a candidate item $i$ as:
\begin{equation}
\label{eq:joint-score}
r_{u,i}=\max_{1\le k\le K}\ \pi_k^u\,(\mathbf{v}_k^u)^\top\mathbf{e}_i,
\end{equation}
where $\pi_k^u$ rescales the score of each interest channel so that scores across interests are placed on a comparable basis. At inference, Eq.~\eqref{eq:joint-score} serves as the ranking function. Since $\pi_k^u\,(\mathbf{v}_k^u)^\top\mathbf{e}_i=(\pi_k^u\mathbf{v}_k^u)^\top\mathbf{e}_i$, the joint score is served by standard per-interest top-$N$ inner-product retrieval with the scaled interests $\pi_k^u\mathbf{v}_k^u$, followed by score-based merging, requiring no change to the existing retrieval pipeline.

\subsection{Optimization}
\label{sec:optimization}
MIMA is trained with two complementary objectives: an interest-item loss $\mathcal{L}_{\mathrm{II}}$ that shapes the representations of interests and items, and a user-interest loss $\mathcal{L}_{\mathrm{UI}}$ that supervises the routing distribution:
\begin{equation}
\label{eq:total-loss}
\mathcal{L}=\mathcal{L}_{\mathrm{II}}+\mathcal{L}_{\mathrm{UI}}.
\end{equation}

Let $\mathcal{P}_u=\{j\mid m_{u,j}=1\}$ denote the index set of valid positives of a training instance. We first construct a negative pool $\mathcal{N}$ following the strategy in \cite{xie2023rethinking}. Then, for each assigned pair $(\bar{\mathbf{v}}_j^u,y_j)$ obtained from Eq.~\eqref{eq:assigned-interest}, we adopt a sampled softmax objective:
\begin{equation}
\label{eq:interest-item-loss}
\mathcal{L}_{\mathrm{II}}=-\frac{1}{|\mathcal{P}_u|}\sum_{j\in\mathcal{P}_u}\log\frac{\exp\!\left((\bar{\mathbf{v}}_j^u)^\top\mathbf{e}_{y_j}\right)}{\exp\!\left((\bar{\mathbf{v}}_j^u)^\top\mathbf{e}_{y_j}\right)+\sum_{i\in\mathcal{N}}\exp\!\left((\bar{\mathbf{v}}_j^u)^\top\mathbf{e}_i\right)}.
\end{equation}

To supervise $\boldsymbol{\pi}_u$, we construct a hinge objective following the joint score defined in Eq.~\eqref{eq:joint-score}, requiring the calibrated score of each positive to exceed that of the hardest negative by a margin $\gamma$. Since raw inner products are unbounded and make a fixed margin scale-sensitive, the similarity terms are computed with the L2-normalized interest and item vectors $\hat{\mathbf{v}}_k^u=\mathbf{v}_k^u/\lVert\mathbf{v}_k^u\rVert$ and $\hat{\mathbf{e}}_i=\mathbf{e}_i/\lVert\mathbf{e}_i\rVert$:
\begin{equation}
\label{eq:user-interest-loss}
\begin{gathered}
r_j^{+}=\max_{1\le k\le K}\ \pi_k^u\operatorname{sg}\!\left[(\hat{\mathbf{v}}_k^u)^\top\hat{\mathbf{e}}_{y_j}\right],\\
r^{-}=\max_{1\le k\le K,\,i\in\mathcal{N}}\ \pi_k^u\operatorname{sg}\!\left[(\hat{\mathbf{v}}_k^u)^\top\hat{\mathbf{e}}_i\right],\\
\mathcal{L}_{\mathrm{UI}}=\frac{1}{|\mathcal{P}_u|}\sum_{j\in\mathcal{P}_u}\max\left(0,\ \gamma-r_j^{+}+r^{-}\right).
\end{gathered}
\end{equation}

Complementing the detached routing input in Eq.~\eqref{eq:routing}, the stop-gradient here on the similarity terms closes the remaining gradient path of $\mathcal{L}_{\mathrm{UI}}$, so it updates only the routing MLP while $\mathcal{L}_{\mathrm{II}}$ exclusively shapes the interest and item representations, keeping the two objectives decoupled in parameter space. We adopt the hinge form because it is structurally consistent with the max-based ranking in Eq.~\eqref{eq:joint-score}, requiring only that the best interest channel of a positive outscore negatives. Since serving follows the same structure, training and inference remain consistent. The overall training and serving procedure of MIMA is summarized in Algorithm~\ref{alg:MIMA}.

\begin{algorithm}[t]
\caption{Training and Inference of MIMA}\label{alg:MIMA}
\KwIn{User interaction sequences $s_u$; interest number $K$.}
\KwOut{Top-$N$ retrieved items for user $u$.}
\BlankLine
Initialize model parameters and queries $\mathbf{Q}\in\mathbb{R}^{(K+1)\times d}$\;
Construct multi-positive training instances $(s_u, \mathcal{Y}_u)$\;

\ForEach{mini-batch $\mathcal{B}$}{
  \ForEach{instance $(s_u, \mathcal{Y}_u) \in \mathcal{B}$}{

    $\mathbf{X}_u \leftarrow [\mathbf{e}_{i_1^u}+\mathbf{p}_1,\ldots,\mathbf{e}_{i_n^u}+\mathbf{p}_n]^\top$\;
    $\mathbf{H}_u, \mathbf{A}_u \leftarrow \operatorname{CausalDecoder}(\mathbf{Q}, \mathbf{X}_u)$\;
    \For{$k = 1$ \KwTo $K$}{
      $\mathbf{v}_k^u \leftarrow \sum_{m=1}^{n} A_{k,m}^u\, \mathbf{e}_{i_m^u}$\tcp*{interest vectors}
    }

    $C_{k,j}^u \leftarrow (\mathbf{v}_k^u)^\top \mathbf{e}_{y_j},\quad \forall\, 1\le k\le K,\, {y_j}\in\mathcal{Y}_u$\;
    $\mathbf{Z}^* \leftarrow \operatorname{Hungarian}(-\mathbf{C}_u)$\tcp*{assignment matrix}
    \ForEach{positive item ${y_j} \in \mathcal{Y}_u$}{
      $\bar{\mathbf{v}}_j^u \leftarrow \sum_{k=1}^{K} Z^*_{k,j}\, \mathbf{v}_k^u$\tcp*{assigned interest}
    }

    $\boldsymbol{\pi}_u \leftarrow \operatorname{softmax}\bigl(\mathbf{W}_2\,\sigma(\mathbf{W}_1\, \operatorname{sg}[\mathbf{h}_{K+1}^u])\bigr)$\;

    Compute $\mathcal{L}_{\mathrm{II}}$ by Eq.~\eqref{eq:interest-item-loss}\;
    Compute $\mathcal{L}_{\mathrm{UI}}$ by Eq.~\eqref{eq:user-interest-loss}\;
  }
  Update model parameters by minimizing $\mathcal{L} = \mathcal{L}_{\mathrm{II}} + \mathcal{L}_{\mathrm{UI}}$\;
}
\BlankLine
\ForEach{user $u$ at inference}{
  Generate $\{\mathbf{v}_k^u\}_{k=1}^{K}$ and $\boldsymbol{\pi}_u$ from $s_u$\;
  Retrieve top-$N$ items ranked by $r_{u,i} = \max_{k}\, \pi_k^u\, (\mathbf{v}_k^u)^\top \mathbf{e}_i$\;
}
\end{algorithm}

\section{Experiments}
\subsection{Experimental Settings}
\subsubsection{\textbf{Datasets}}
\label{sec:datasets}
We conduct experiments on three public datasets and an industrial dataset to evaluate our method. For public datasets, we use \textbf{Books} and \textbf{Beauty} \cite{ni2019justifying}, which contain user review records from the Amazon platform, and \textbf{Gowalla} \cite{cho2011friendship}, a location-based social network with user check-in records. Following standard practice \cite{cen2020controllable, xie2023rethinking}, we treat all interactions as implicit feedback, remove users and items with fewer than 5 interactions, and sort user behaviors chronologically. We set the maximum sequence length to 20 for Books and Beauty, and 40 for Gowalla. For the industrial dataset, we collect \textbf{Industry} from the international e-commerce platform Lazada, specifically from its Thailand marketplace. The dataset spans 11 days with a maximum sequence length of 1024. 
The statistics of all four datasets are summarized in Table~\ref{tab:dataset}.

For multi-positive target construction, positive sets on Industry are built directly from real serving requests as described in Section~\ref{sec:multi-positive}. In contrast, the public datasets provide no request logs, and we thus approximate request-level co-occurrence with time windows. Specifically, we partition each user's chronological interactions into consecutive one-day windows, where one day is the finest timestamp granularity in Books and Beauty and is also adopted for Gowalla for consistency. Items falling into the same window are grouped into a positive set.

\begin{table}[t]
\centering
\caption{Statistics of the experimental datasets.}
\label{tab:dataset}
\resizebox{0.95\linewidth}{!}{
\begin{tabular}{l|cccc}
\toprule
Dataset & \#Users & \#Items & \#Interactions & Density \\
\midrule
Books & 603,668 & 367,982 & 8,898,041 & 0.0040\% \\
Beauty & 40,226 & 67,345 & 353,962 & 0.0131\% \\
Gowalla & 65,506 & 174,605 & 2,061,264 & 0.0180\% \\
Industry & 5,913,083 & 7,604,006 & 63,293,772 & 0.0001\% \\
\bottomrule
\end{tabular}
}
\end{table}

\subsubsection{\textbf{Baselines}} We compare our method against both general single-interest recommendation models and state-of-the-art multi-interest models. The single-interest baselines include:
\begin{itemize}
    \item \textbf{POP}: a simple method that always recommends the most popular items to users.
    \item \textbf{YouTube-DNN} \cite{covington2016deep}: employs a deep neural network to learn user representations from historical interactions.
    \item \textbf{GRU4Rec} \cite{hidasi2015session}: employs Gated Recurrent Units to encode user behavior sequences for sequential recommendation.
\end{itemize}

The multi-interest baselines include:
\begin{itemize}
    \item \textbf{MIND} \cite{li2019multi}: the pioneering multi-interest model that adopts capsule network to extract multiple interest vectors.
    \item \textbf{ComiRec} \cite{cen2020controllable}: introduces multi-head self-attention for interest extraction and a diversity-aware aggregation module for controllable multi-interest recommendation.
    \item \textbf{PIMI} \cite{chen2021exploring}: incorporates time periodicity and item interactivity into multi-interest sequential recommendation.
    \item \textbf{RE4} \cite{zhang2022re4}: applies re-contrast, re-attend, and re-construct regularization to improve interest representation quality.
    \item \textbf{REMI} \cite{xie2023rethinking}: a training framework that introduces interest-aware hard negative mining and routing regularization to address easy negatives and routing collapse.
    \item \textbf{DisMIR} \cite{du2024disentangled}: uses spectral clustering on the co-occurrence graph for disentangled multi-interest representations.
    \item \textbf{NPRec} \cite{jiang2025auto}: models the distribution over user preference via neural processes, providing uncertainty estimation.
\end{itemize}

\begin{table*}[t]
\centering
\caption{Performance comparison between MIMA and baselines on three public datasets. R@$N$, N@$N$, and H@$N$ refer to Recall@$N$, NDCG@$N$, and HR@$N$, respectively. The best results are shown in \textbf{bold} and the second-best results are \underline{underlined}. All improvements of MIMA over the best-performing baseline are statistically significant with $p < 0.01$ based on paired t-tests.}
\label{tab:performance}
    \renewcommand{\arraystretch}{0.95} 
    \setlength{\tabcolsep}{4pt}
\resizebox{0.95\textwidth}{!}{
\begin{tabular}{ll|ccc|ccccccc|cc}
\toprule
\multirow{2}{*}{Dataset} & \multirow{2}{*}{Metric} & \multicolumn{3}{c|}{Single-Interest Methods} & \multicolumn{7}{c|}{Multi-Interest Methods} & \multicolumn{2}{c}{Ours} \\ \cmidrule(lr){3-5} \cmidrule(lr){6-12} \cmidrule(lr){13-14}
 &  & POP & Y-DNN & GRU4Rec & MIND & ComiRec & PIMI & RE4 & REMI & DisMIR & NPRec & MIMA & Improv. \\ \midrule
\multirow{6}{*}{Books} 
 & R@20 & 0.0159 & 0.0465 & 0.0499 & 0.0421 & 0.0534 & 0.0531 & 0.0515 & 0.0825 & \underline{0.0883} & 0.0604 & \textbf{0.1017} & 15.18\% \\
 & N@20 & 0.0143 & 0.0391 & 0.0442 & 0.0333 & 0.0416 & 0.0409 & 0.0396 & 0.0620 & \underline{0.0671} & 0.0502 & \textbf{0.0743} & 10.73\% \\
 & H@20 & 0.0345 & 0.1045 & 0.1164 & 0.0894 & 0.1113 & 0.1103 & 0.1068 & 0.1645 & \underline{0.1792} & 0.1334 & \textbf{0.1960} & 9.38\% \\
 & R@50 & 0.0281 & 0.0736 & 0.0800 & 0.0661 & 0.0848 & 0.0848 & 0.0822 & 0.1196 & \underline{0.1355} & 0.1003 & \textbf{0.1568} & 15.72\% \\
 & N@50 & 0.0193 & 0.0452 & 0.0509 & 0.0385 & 0.0487 & 0.0485 & 0.0474 & 0.0657 & \underline{0.0744} & 0.0595 & \textbf{0.0833} & 11.96\% \\
 & H@50 & 0.0601 & 0.1594 & 0.1769 & 0.1365 & 0.1718 & 0.1735 & 0.1686 & 0.2303 & \underline{0.2621} & 0.2090 & \textbf{0.2901} & 10.68\% \\ \midrule
\multirow{6}{*}{Beauty} 
 & R@20 & 0.0176 & 0.0295 & 0.0257 & 0.0405 & 0.0296 & 0.0268 & 0.0299 & 0.0494 & \underline{0.0504} & 0.0354 & \textbf{0.0540} & 7.14\% \\
 & N@20 & 0.0129 & 0.0215 & 0.0189 & 0.0275 & 0.0198 & 0.0181 & 0.0190 & \underline{0.0335} & 0.0333 & 0.0264 & \textbf{0.0374} & 11.64\% \\
 & H@20 & 0.0336 & 0.0554 & 0.0490 & 0.0753 & 0.0534 & 0.0500 & 0.0539 & 0.0890 & \underline{0.0895} & 0.0689 & \textbf{0.0967} & 8.04\% \\
 & R@50 & 0.0377 & 0.0509 & 0.0453 & 0.0667 & 0.0475 & 0.0461 & 0.0518 & 0.0807 & \underline{0.0841} & 0.0612 & \textbf{0.0881} & 4.76\% \\
 & N@50 & 0.0200 & 0.0269 & 0.0229 & 0.0327 & 0.0241 & 0.0223 & 0.0262 & 0.0392 & \underline{0.0399} & 0.0306 & \textbf{0.0423} & 6.02\% \\
 & H@50 & 0.0696 & 0.0890 & 0.0788 & 0.1171 & 0.0863 & 0.0803 & 0.0964 & 0.1397 & \underline{0.1404} & 0.1049 & \textbf{0.1481} & 5.48\% \\ \midrule
\multirow{6}{*}{Gowalla} 
 & R@20 & 0.0228 & 0.0884 & 0.0787 & 0.0874 & 0.0630 & 0.0594 & 0.0718 & \underline{0.1328} & 0.1239 & 0.0904 & \textbf{0.1378} & 3.77\% \\
 & N@20 & 0.0483 & 0.1419 & 0.1286 & 0.1325 & 0.0971 & 0.0932 & 0.1116 & \underline{0.1774} & 0.1729 & 0.1314 & \textbf{0.1828} & 3.04\% \\
 & H@20 & 0.1114 & 0.3244 & 0.3025 & 0.3088 & 0.2336 & 0.2258 & 0.2650 & \underline{0.4151} & 0.4036 & 0.3135 & \textbf{0.4320} & 4.07\% \\
 & R@50 & 0.0366 & 0.1447 & 0.1288 & 0.1475 & 0.1172 & 0.1079 & 0.1310 & \underline{0.2056} & 0.2018 & 0.1510 & \textbf{0.2205} & 7.25\% \\
 & N@50 & 0.0574 & 0.1477 & 0.1368 & 0.1447 & 0.1208 & 0.1113 & 0.1314 & 0.1776 & \underline{0.1790} & 0.1439 & \textbf{0.1878} & 4.92\% \\
 & H@50 & 0.1589 & 0.4471 & 0.4210 & 0.4474 & 0.3793 & 0.3555 & 0.4097 & 0.5453 & \underline{0.5520} & 0.4488 & \textbf{0.5730} & 3.80\% \\
\bottomrule
\end{tabular}
}
\end{table*}

\subsubsection{\textbf{Evaluation Protocols}}
Following \cite{cen2020controllable, xie2023rethinking}, for the three public datasets, we split users into training, validation, and test sets with a ratio of 8:1:1. For evaluation, we use the first 80\% of the chronological behavior sequence to infer interest representations, and compute metrics on the remaining 20\%. We report Recall@$N$, NDCG@$N$ (Normalized Discounted Cumulative Gain), and HR@$N$ (Hit Rate) with $N \in \{20, 50\}$ as evaluation metrics. For the Industry dataset, we use the first 10 days of logs for training and the final day for testing, and adopt HR@$N$ with $N \in \{100, 500, 1000\}$ as the metric to accommodate the substantially larger candidate pool.

\subsubsection{\textbf{Implementation Details}}
We implement MIMA and all baselines using PyTorch. For all methods, the embedding dimension $d$ is set to 64 and the batch size is set to 128. The size of the negative pool $\mathcal{N}$ is set to 1280 following \cite{xie2023rethinking}. For MIMA, the decoder consists of 2 layers and the number of attention heads is set to 2 on Books and Gowalla and 8 on Beauty. The number of interests $K$ (which also determines the positive set capacity) is set to 4 on Gowalla and 8 on Books and Beauty. The hinge margin $\gamma$ is set to 0.02 on all datasets. We train MIMA using a learning rate of $1\times10^{-3}$ on Books and $5\times10^{-4}$ on Gowalla and Beauty.

\subsection{Overall Performance Comparison}
Table~\ref{tab:performance} compares MIMA with single- and multi-interest baselines on three public datasets. Key findings are: \textbf{(i) MIMA consistently achieves the best performance across all datasets,} outperforming the second-best method by 3.04\%-15.72\%, with all improvements statistically significant at \(p<0.01\). These results show the overall effectiveness of MIMA and confirm the value of applying the multi-positive paradigm to multi-interest recommendation.
\textbf{(ii) Multi-interest modeling does not always lead to better performance.} Early methods such as MIND, ComiRec, and PIMI sometimes underperform single-interest baselines. A possible reason is that their extracted interests collapse into redundant vectors, and the extra interest capacity brings little benefit while fragmenting the supervision signal. In contrast, later methods such as REMI and DisMIR alleviate this collapse problem and perform better. This confirms interest collapse as a key bottleneck, while MIMA addresses it directly at the supervision level rather than through auxiliary regularization.
\textbf{(iii) MIMA shows a clearer advantage on sparse datasets.} The relative improvement grows as interaction density decreases, from 3.04\%–7.25\% on the densest Gowalla to 4.76\%–11.64\% on Beauty and 9.38\%–15.72\% on the sparsest Books. This suggests that the multi-positive supervision of MIMA provides richer and more reliable learning signals when observed interactions are limited, thereby enhancing its effectiveness under data sparsity.

\subsection{Ablation Study}
To verify the contribution of key components in MIMA, we conduct an ablation study and compare the full MIMA against three variants: \textbf{w/o MP (Multi-Positive)} removes the multi-positive supervision together with the corresponding exclusive interest assignment, degenerating into the conventional single-positive training paradigm. \textbf{w/o DC (Decoder)} replaces the causal decoder with the widely used ComiRec-style attention extractor for interest generation. \textbf{w/o RT (Routing)} removes the user-interest routing module and treats all interest channels equally when merging their scores.

As shown in Table~\ref{tab:ablation}, \textbf{w/o MP} suffers the largest performance drop across all datasets, confirming that multi-positive supervision with exclusive assignment is the core of MIMA, as it converts concurrent behaviors into differentiated training signals and alleviates interest collapse through the training objective itself. \textbf{w/o RT} also degrades performance consistently, suggesting that scores from different interest channels are not directly comparable, and calibrating them with the estimated activation strength yields a more reliable merged ranking. Finally, \textbf{w/o DC} underperforms the full model, indicating that the causal decoder, where each interest query attends to previously generated ones, produces more complementary interests than independent attention-based extraction.

\begin{table}[t]
\centering
\caption{Ablation study on three public datasets.}
\label{tab:ablation}
\renewcommand{\arraystretch}{0.98}
\setlength{\tabcolsep}{2.5pt}
\resizebox{\linewidth}{!}{
\begin{tabular}{lc|ccc|ccc}
\toprule
Dataset & Variant & R@20 & N@20 & H@20 & R@50 & N@50 & H@50 \\ \midrule
\multirow{4}{*}{Books}
 & w/o MP & 0.0871 & 0.0653 & 0.1716 & 0.1256 & 0.0691 & 0.2399 \\
 & w/o DC & 0.0985 & 0.0716 & 0.1905 & 0.1496 & 0.0795 & 0.2781 \\
 & w/o RT & 0.0973 & 0.0711 & 0.1879 & 0.1481 & 0.0787 & 0.2748 \\
 & Full & \textbf{0.1017} & \textbf{0.0743} & \textbf{0.1960} & \textbf{0.1568} & \textbf{0.0833} & \textbf{0.2901} \\ \midrule
\multirow{4}{*}{Beauty}
 & w/o MP & 0.0478 & 0.0326 & 0.0880 & 0.0775 & 0.0377 & 0.1352 \\
 & w/o DC & 0.0511 & 0.0358 & 0.0930 & 0.0850 & 0.0417 & 0.1472 \\
 & w/o RT & 0.0498 & 0.0343 & 0.0900 & 0.0834 & 0.0405 & 0.1414 \\
 & Full & \textbf{0.0540} & \textbf{0.0374} & \textbf{0.0967} & \textbf{0.0881} & \textbf{0.0423} & \textbf{0.1481} \\ \midrule
\multirow{4}{*}{Gowalla}
 & w/o MP & 0.1298 & 0.1672 & 0.3993 & 0.1932 & 0.1712 & 0.5118 \\
 & w/o DC & 0.1366 & 0.1778 & 0.4238 & 0.2178 & 0.1839 & 0.5627 \\
 & w/o RT & 0.1324 & 0.1788 & 0.4192 & 0.2173 & 0.1861 & 0.5663 \\
 & Full & \textbf{0.1378} & \textbf{0.1828} & \textbf{0.4320} & \textbf{0.2205} & \textbf{0.1878} & \textbf{0.5730} \\
\bottomrule
\end{tabular}
}
\end{table}

\subsection{Advanced Analysis}

\subsubsection{\textbf{Analysis of Different Interest Assignment Methods}}
\label{sec:assignment-analysis}
To evaluate the effectiveness of our interest assignment strategy, we compare the Hungarian assignment adopted in MIMA with two alternatives: greedy assignment and Sinkhorn-based soft assignment. As shown in Table~\ref{tab:assignment}, Hungarian achieves the best performance across the three datasets, which is consistent with the fact that it exactly solves the linear assignment problem and maximizes the total similarity of all positive-interest pairs given the current interest vectors. Greedy is a strong competitor, as it also performs hard one-to-one matching and keeps the gradient of each positive concentrated on a single interest. However, its step-wise local decisions may lock in suboptimal pairs, making it slightly inferior to Hungarian overall. Sinkhorn lags behind, especially on the sparse Books dataset, because its soft matching spreads the gradient of each positive over multiple interests, weakening interest exclusivity and diluting the supervision signal. These results suggest that exact and exclusive assignment benefits multi-interest specialization.

\begin{table}[t]
\centering
\caption{Performance comparison of different interest assignment methods on the three public datasets.}
\label{tab:assignment}
\renewcommand{\arraystretch}{1}
\setlength{\tabcolsep}{2.5pt}
\resizebox{\linewidth}{!}{
\begin{tabular}{ll|ccc|ccc}
\toprule
Dataset & Method & R@20 & N@20 & H@20 & R@50 & N@50 & H@50 \\ \midrule
\multirow{3}{*}{Books}
 & Greedy   & 0.1004          & 0.0736          & 0.1955          & 0.1515          & 0.0810          & 0.2820          \\
 & Sinkhorn & 0.0892          & 0.0656          & 0.1768          & 0.1325          & 0.0716          & 0.2515          \\
 & Hungarian  & \textbf{0.1017} & \textbf{0.0743} & \textbf{0.1960} & \textbf{0.1568} & \textbf{0.0833} & \textbf{0.2901} \\ \midrule
\multirow{3}{*}{Beauty}
 & Greedy   & 0.0518          & 0.0364          & 0.0920          & 0.0877          & 0.0421          & 0.1462          \\
 & Sinkhorn & 0.0509          & 0.0351          & 0.0915          & 0.0872          & 0.0421          & 0.1459          \\
 & Hungarian  & \textbf{0.0540} & \textbf{0.0374} & \textbf{0.0967} & \textbf{0.0881} & \textbf{0.0423} & \textbf{0.1481} \\ \midrule
\multirow{3}{*}{Gowalla}
 & Greedy   & 0.1376          & 0.1818          & 0.4282          & 0.2187          & 0.1860          & 0.5654          \\
 & Sinkhorn & 0.1377          & 0.1826          & 0.4312          & 0.2202          & 0.1874          & 0.5724          \\
 & Hungarian  & \textbf{0.1378} & \textbf{0.1828} & \textbf{0.4320} & \textbf{0.2205} & \textbf{0.1878} & \textbf{0.5730} \\
\bottomrule
\end{tabular}
}
\end{table}

\subsubsection{\textbf{Visualization of Interest Representations}}
\begin{figure}[t]
    \centering
    \includegraphics[width=\linewidth]{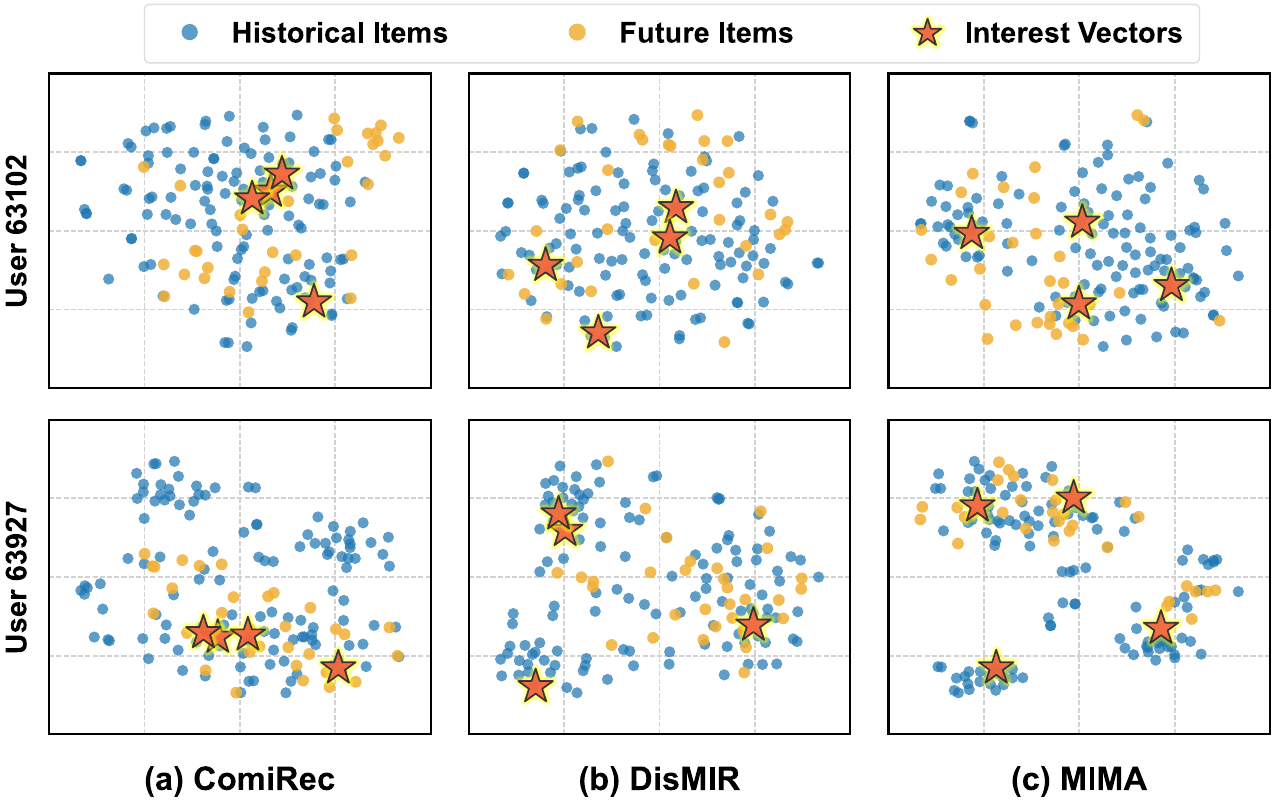}
    \caption{t-SNE visualization of interest representations.}
    \label{fig:tsne}
\end{figure}

We visualize the interest vectors via t-SNE to verify whether MIMA alleviates interest collapse. Specifically, we sample two users with 150 interactions from the Gowalla test set, and project their historical items, future items, and the four interests into a shared space. As shown in Figure~\ref{fig:tsne}, the interests of ComiRec heavily overlap, indicating severe collapse where multiple vectors cover the same behavior region. DisMIR separates the interests to some extent but still exhibits redundancy and leaves certain behaviors uncovered. In contrast, MIMA produces the most differentiated interests that tend to align with distinct behavior clusters and provide better coverage of the regions where future items reside, confirming its effectiveness in capturing diverse user interests and mitigating collapse.

\subsubsection{\textbf{Effect of Number of Interests}}
\begin{figure}[t]
    \centering
    \includegraphics[width=\linewidth]{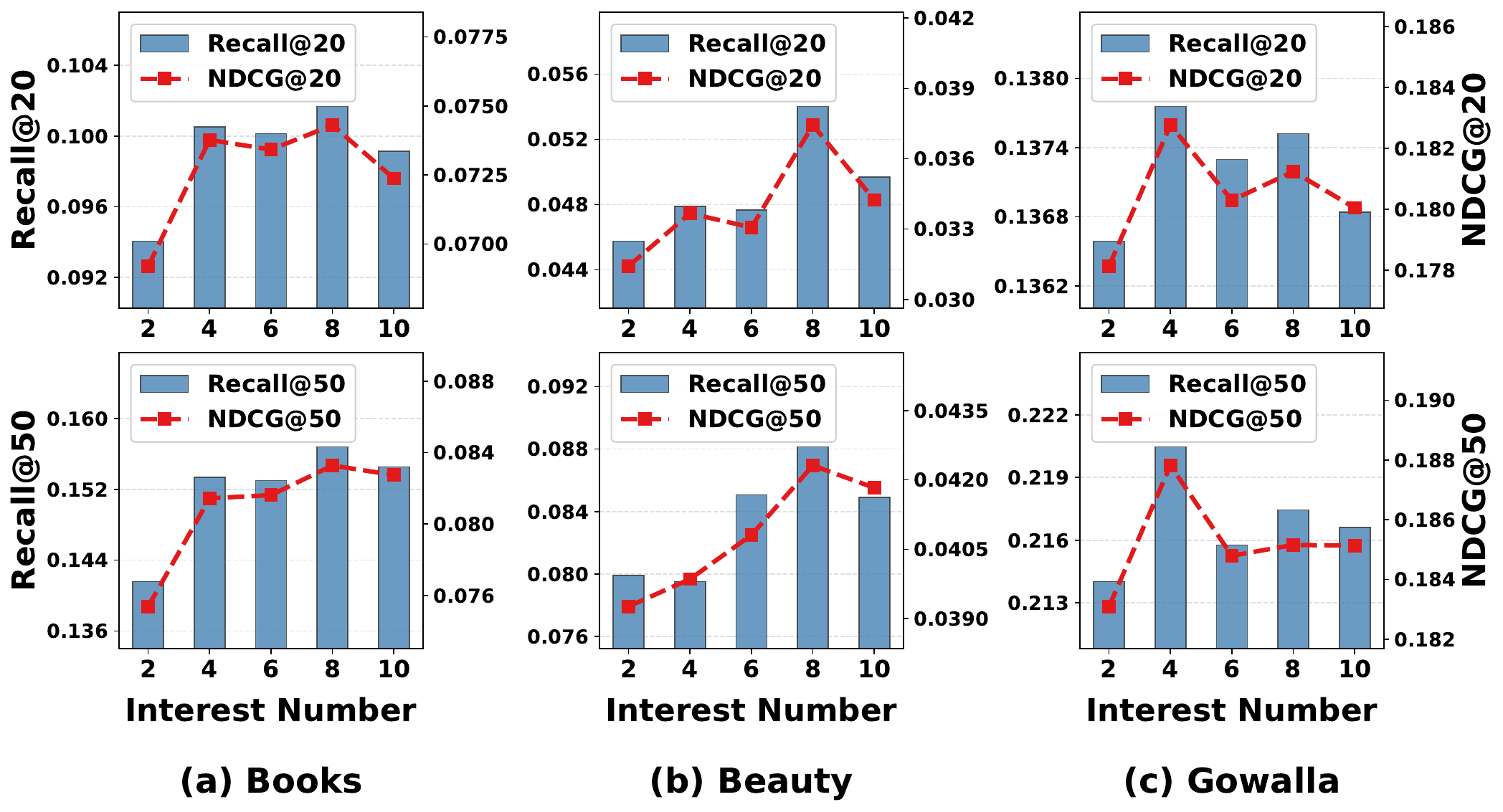}
    \caption{Effect of the number of interests.}
    \label{fig:hyperparameter_k}
\end{figure}

We study the impact of the interest number $K$ by varying it in $\{2,4,6,8,10\}$. Since the number of positives is always set to $K$, varying $K$ also changes the amount of multi-positive supervision. As shown in Figure~\ref{fig:hyperparameter_k}, performance improves substantially when $K$ increases from 2 to 4, since too few interests cannot cover the diverse preferences of users. The best results are achieved at $K=8$ on Books and Beauty and $K=4$ on Gowalla, suggesting that the optimal $K$ depends on the preference diversity of the dataset. When $K$ grows beyond the optimum, performance declines slightly, as redundant interests over-partition user behaviors and introduce noise into retrieval.

\subsubsection{\textbf{Effect of Number of Decoder Layers and Heads}}
\begin{figure}[t]
    \centering
    \includegraphics[width=\linewidth]{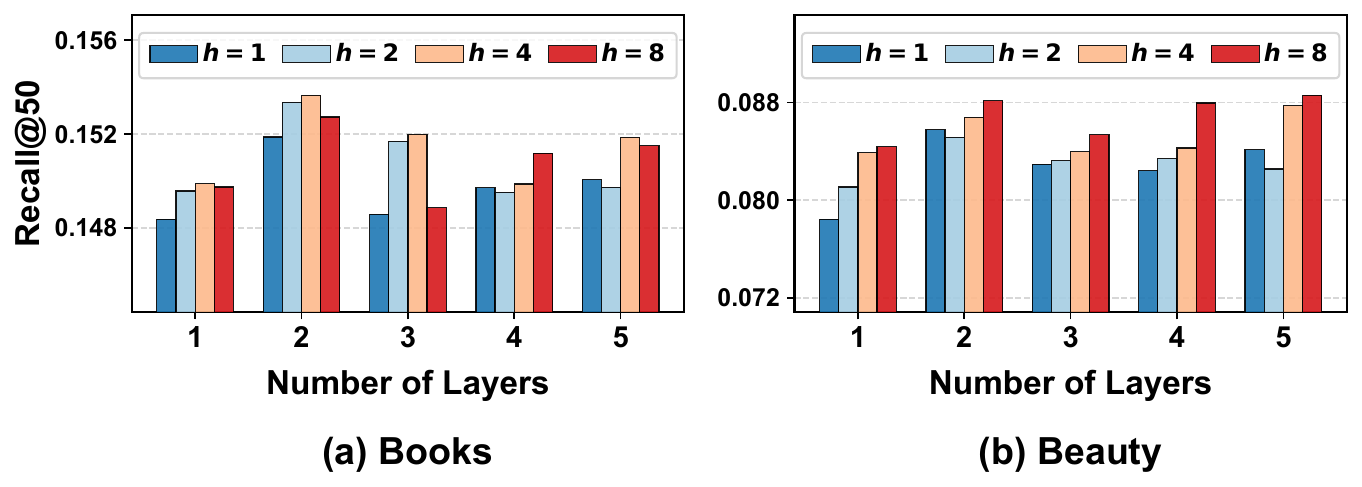}
    \caption{Effect of the number of decoder layers and attention heads on recommendation performance.}
    \label{fig:layer_head_grid_2x2}
\end{figure}

We investigate the sensitivity of MIMA to the number of decoder layers and attention heads $h$ on Books and Beauty. As shown in Figure~\ref{fig:layer_head_grid_2x2}, a two-layer decoder already achieves the best or near-best performance on both datasets, suggesting that a shallow decoder is sufficient for interest generation. Regarding attention heads, the optimal $h$ varies across datasets. Beauty benefits from more heads with $h=8$ performing best, whereas on Books a moderate $h$ of 2 or 4 is sufficient. Accordingly, we adopt a two-layer decoder for all datasets, and set $h=2$ for Books and Gowalla and $h=8$ for Beauty.

\subsection{Industrial Results}
\subsubsection{\textbf{Offline Evaluation}}
To assess the effectiveness of MIMA in industrial scenarios, we conduct an offline evaluation on the Industry dataset, where the number of interests is set to 5 for all multi-interest methods. As shown in Table~\ref{tab:industrial_offline}, multi-interest methods generally outperform single-interest ones, verifying the benefit of modeling diverse user preferences in real-world scenarios. Notably, DisMIR, which leads on the public datasets, falls behind here, likely because its spectral clustering over the item co-occurrence graph degrades when the graph becomes extremely sparse and noisy at industrial scale. MIMA still achieves the best HR under all $N$, consistently outperforming the strongest baseline.

\textbf{Quantitative Analysis of Interest Collapse.}
To further examine whether these recommendation performance gains come with mitigated interest collapse, we propose the Interest Discrimination Margin (IDM@$N$), which measures the similarity margin of each retrieved candidate to its retrieving interest over its strongest rival:
\begin{equation}
\label{eq:idm}
\text{IDM@}N=\frac{1}{|\mathcal{U}|KN}\sum_{u\in\mathcal{U}}\sum_{k=1}^{K}\sum_{i\in\Omega^{(k)}_u}\!\Big((\hat{\mathbf{v}}_k^u)^\top\hat{\mathbf{e}}_i-\max_{j\neq k}\,(\hat{\mathbf{v}}_j^u)^\top\hat{\mathbf{e}}_i\Big),
\end{equation}
where $\Omega^{(k)}_u$ denotes the top-$N$ items retrieved by the $k$-th interest of user $u$. A higher IDM indicates that the candidates retrieved by each interest align distinctly with that interest rather than with the others, implying less collapse. We report IDM for all multi-interest methods, among which early ones such as MIND and PIMI suffer the most severe collapse, while those with explicit anti-collapse designs such as REMI and DisMIR obtain notably higher IDM. This trend is consistent with the HR results, corroborating that interest collapse is a key bottleneck in industrial retrieval. MIMA achieves the highest IDM, confirming that its exclusive multi-positive supervision yields genuinely differentiated interests at industrial scale.

\begin{table}[t]
\centering
\caption{Offline results on the industrial dataset. IDM@$N$ quantitatively measures the degree of interest collapse.}
\label{tab:industrial_offline}
\renewcommand{\arraystretch}{1.0}
\setlength{\tabcolsep}{3pt}
\resizebox{\linewidth}{!}{
\begin{tabular}{ll|ccc|ccc}
\toprule
\multirow{2}{*}{Category} & \multirow{2}{*}{Method} & \multicolumn{3}{c|}{HR@$N$} & \multicolumn{3}{c}{IDM@$N$} \\ \cmidrule(lr){3-5} \cmidrule(lr){6-8}
 & & 100 & 500 & 1000 & 100 & 500 & 1000 \\ \midrule
\multirow{2}{*}{\shortstack{Single-Int.}} 
 & Y-DNN & 0.0702 & 0.1497 & 0.1986 & -- & -- & -- \\
 & GRU4Rec & 0.1904 & 0.3195 & 0.3806 & -- & -- & -- \\  \midrule
\multirow{7}{*}{\shortstack{Multi-Int.}}
 & MIND & 0.1863 & 0.3270 & 0.3988 & 0.0510 & 0.0416 & 0.0331 \\
 & ComiRec & 0.2105 & 0.3615 & 0.4353 & 0.0911 & 0.0734 & 0.0654 \\
 & PIMI & 0.1234 & 0.2412 & 0.3081 & 0.0500 & 0.0403 & 0.0357 \\
 & RE4 & 0.1950 & 0.3356 & 0.4067 & 0.0732 & 0.0557 & 0.0480 \\
 & REMI & \underline{0.2119} & \underline{0.3645} & \underline{0.4390} & \underline{0.1227} & \underline{0.1010} & \underline{0.0879} \\
 & DisMIR & 0.1871 & 0.3303 & 0.4040 & 0.0951 & 0.0759 & 0.0669 \\
 & NPRec & 0.1782 & 0.3058 & 0.3704 & 0.0722 & 0.0550 & 0.0460 \\  \midrule
\multirow{2}{*}{Ours}
 & MIMA & \textbf{0.2288} & \textbf{0.4091} & \textbf{0.4962} & \textbf{0.1453} & \textbf{0.1207} & \textbf{0.1087} \\ 
 & Improv. & 7.98\% & 12.24\% & 13.03\% & 18.42\% & 19.50\% & 23.66\% \\
\bottomrule
\end{tabular}
}
\end{table}

\begin{table}[t]
\centering
\caption{Online A/B testing results.}
\label{tab:online_ab}
\renewcommand{\arraystretch}{1}
\small
\begin{tabular}{ccc}
\toprule
Metric & Transaction Count & Transaction Amount\\ \midrule
Relative Improv. & +5.60\% & +5.44\%\\
\bottomrule
\end{tabular}
\end{table}

\subsubsection{\textbf{Online A/B Test}}
We deploy MIMA in the retrieval stage of homepage recommendation on a large-scale e-commerce platform, replacing the existing ComiRec-like matching model, and conduct a 7-day A/B test. As shown in Table~\ref{tab:online_ab}, MIMA achieves relative improvements of 5.60\% in transaction count and 5.44\% in transaction amount. Moreover, its exclusive impression ratio, i.e., the proportion of impressions contributed by items recalled solely by this channel, increases by 5.12 percentage points, indicating stronger complementary recall beyond existing retrieval channels.

\subsubsection{\textbf{Case Study}}
\begin{figure}[t]
    \centering
    \includegraphics[width=0.95\linewidth]{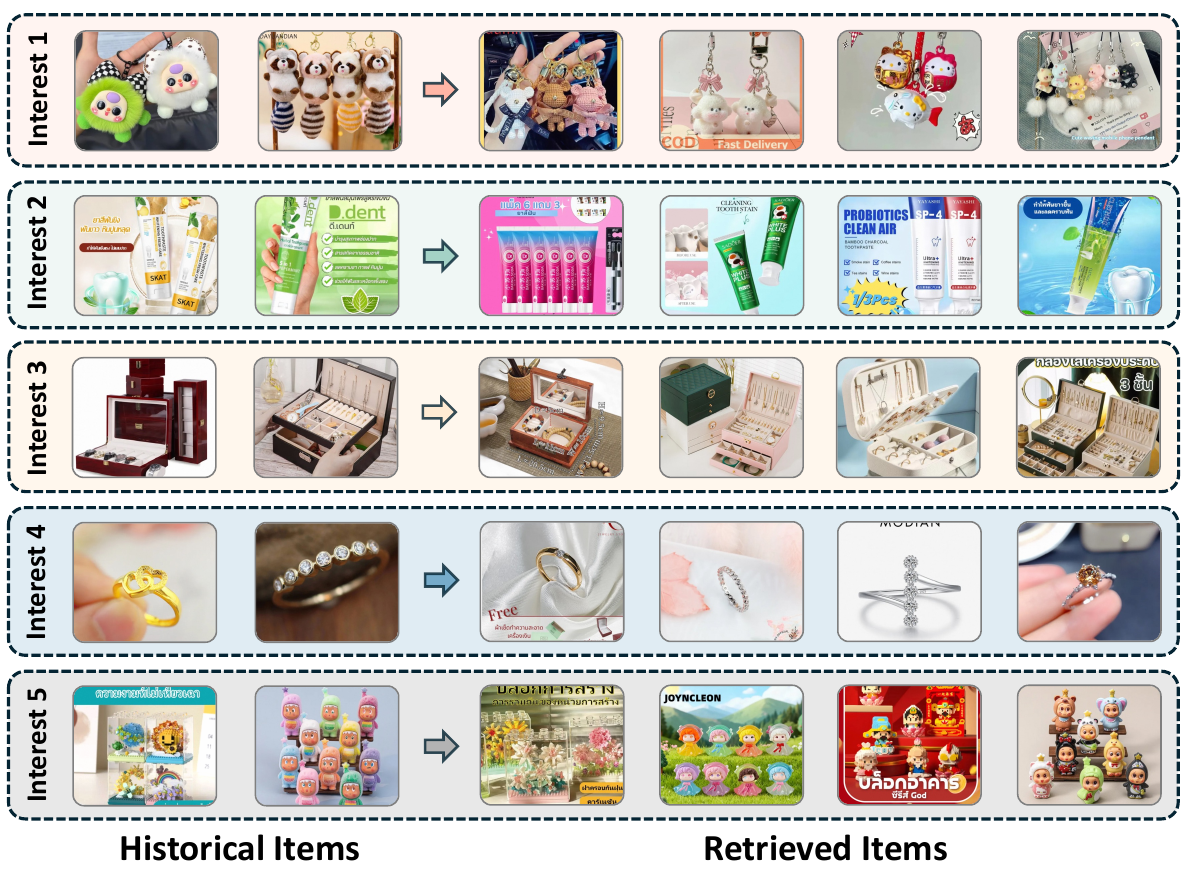}
    \caption{Case study of a sampled user.}
    \label{fig:case_study}
\end{figure}

Beyond quantitative metrics, we further examine whether each interest captures a meaningful and distinguishable preference. We sample a user from the industrial dataset and visualize the five learned interests in Figure~\ref{fig:case_study}, where each row shows the historical items associated with one interest on the left and the items retrieved by the corresponding interest vector on the right. The five interests clearly correspond to five distinct categories with a clear division of labor and no overlap. Moreover, the items retrieved by each interest are highly consistent with the corresponding historical preference, verifying the differentiation and interpretability of the interest representations learned by MIMA.

\section{Conclusion}
In this paper, we revisit the supervision paradigm of multi-interest recommendation and highlight single-positive training as one important factor behind interest collapse. We propose MIMA, a framework based on multi-positive exclusive assignment. By grouping concurrent behaviors into positive sets, generating complementary interests with a causal decoder, and enforcing exclusive one-to-one matching between interests and positives, MIMA lets interest differentiation emerge directly from the training objective rather than auxiliary regularization. A lightweight user-interest routing module further estimates the activation strength of each interest, yielding calibrated and comparable scores across interest channels at inference. Extensive offline experiments on three public datasets and a large-scale industrial dataset, together with an online A/B test, demonstrate that MIMA consistently outperforms state-of-the-art baselines and brings significant business gains.


\balance
\bibliographystyle{ACM-Reference-Format}
\bibliography{sample-base}

\end{document}